\documentclass[aps,prd,groupedaddress]
{revtex4-2}

\usepackage[colorlinks,urlcolor=blue, citecolor=blue,linkcolor=blue]{hyperref}
\usepackage{verbatim} 
\usepackage{graphicx}
\usepackage{amsmath}
\usepackage{xcolor}

\usepackage{lineno}

\begin{document}
\title{Resonant Parameters of Vector Charmonium-like States above 4.4 GeV}

\author{Chunhua Li$^{1,2}$~\footnote {E-mail: chunhua@njnu.edu.cn}, Wanting Liu$^{1}$, Linfa Tang$^{1}$, Ying Ding$^2$\\
$^1$Department of physics and electronic technology,
Liaoning Normal University, 
Dalian, 116029, P.R.China\\
$^2$School of physics and technology,
Nanjing Normal University, 
Nanjing, 210023, P.R.China
}


\begin{abstract}
We analysis the $\sqrt{s}$-dependent line shapes of the $e^+e^-\to D_s^{+}D_{s1}^{*-}(2536)$, $D_s^{+}D_{s2}^{*-}(2573)$, $\phi\chi_{c1,2}$, $K^+K^-J/\psi$, $K_S^0 K_S^0 J/\psi$, and $K^+K^-\psi(2S)$ cross sections measured by the BESIII experiment using the four resonant structures $\psi(4230)$, $\psi(4500)$, $\psi(4660)$, and $\psi(4710)$, by performing a simultaneous $\chi^2$-minimized fit. Their masses and widths are obtained. We find that the processes $e^+e^-\to D_s^{+}D_{s1}^{*-}(2536)$, $e^+e^-\to D_s^{+}D_{s2}^{*-}(2573)$, and $e^+e^-\to \phi\chi_{c1,2}$ are all dominantly produced via the $\psi(4660)$ and $\psi(4710)$ decays.
\end{abstract}

\maketitle

\section{Introduction}

The observation of the $X(3872)$ in 2003~\cite{Belle:2003nnu}, the first quarkonium-like state discovered by the Belle experiment, ushered in a new era for the study of exotic states beyond the conventional quarkonium spectrum. Since then, numerous quarkonium-like states have been reported by various experiments, including Belle, BaBar, LHCb, and BESIII.
The potentially exotic nature often emerges when the observed states cannot be fitted into the theoretically predicted spectrum based on potential models.
Among the reported charmonium-like states, vector charmonium-like states with the quantum number $J^{PC}=1^{--}$ represent one of the most interesting and attractive topics.
The $\psi(4230)$ is the first vector charmonium-like state reported by BaBar via the initial-state radiation process $e^+e^-\to\gamma\pi\pi J/\psi$~\cite{BaBar:2005hhc}.

\section{Charmonium spectrum}

The charmonium spectrum can be 
well described by the potential quark models.
Ref.~\cite{Barnes:2005pb} calculated the charmonium spectrum 
with the nonrelativistic and relativistic potential models.
In the nonrelativistic potential model (NI), 
the central potential is 
\begin{equation}
V_0^{(c\bar{c})}(r) = -\frac{4}{3}\frac{\alpha_s}{r} + br+C_0.
\end{equation}
The parameters $b$ and
$\alpha_s$ denote the strength of the confinement and strong coupling of the one-gluon-exchange potential, respectively, and the $C_0$ is zero point energy.
The spin-dependent terms include  spin-spin, spin-orbit, and tensor operators, are expressed as
\begin{equation}
V_{\text{sp}} = \frac{32\pi\alpha_s}{9m_c^2}\tilde{\delta}_\sigma(r)\vec{S}_c \cdot \vec{S}_{\bar{c}}+ \frac{1}{m_c^2}\left(\frac{2\alpha_s}{r^3} - \frac{b}{2r}\right)\vec{L}\cdot\vec{S} + \frac{1}{m_c^2}\frac{4\alpha_s}{r^3}\mathrm{T}.
\end{equation}
where $\tilde{\delta}_\sigma(r) = \left(\sigma/\sqrt{\pi}\right)^3 e^{-\sigma^2 r^2}$.

The Godfrey–Isgur (GI) model extends the nonrelativistic model to a relativistic framework by replacing the quark mass with the quark kinetic energy, incorporating a relativistic dispersion relation for the quark kinetic energy, and introducing a QCD-motivated running coupling $\alpha_s(r)$.

Based on these two models, this paper calculates the charmonium spectrum and further predicts the E1 and M1 electromagnetic transition rates between them, as well as the strong partial and total decay widths for states above the open-charm threshold. These predictions provide valuable theoretical inputs for relevant experimental searches.
Conventional charmonium states below the open-charm threshold, such as $\eta_c(1S)$, $\psi(1S)$, $\psi(2S)$ and $\chi_{c0,1,2}(1P)$, as well as their decay properties, are well described and experimentally verified.

However, the conventional quark model becomes questionable when describing higher charmonium states with masses above the open-charm threshold. 
As pointed by Ref.~\cite{Deng:2023mza}, these discrepancies may arise from quantum fluctuations, i.e., the creation and annihilation of $q\bar{q}$ pairs in the vacuum which can produce significant effects. This is known as the ``coupled-channel effect'' mediated by virtual charm meson loops, which becomes essential for describing certain highly excited charmonium states.
It is found that the masses of the higher excited states are often lowered by the coupled-channel effects.

\begin{figure*}[!h]
\begin{center}
\includegraphics[scale=0.35]{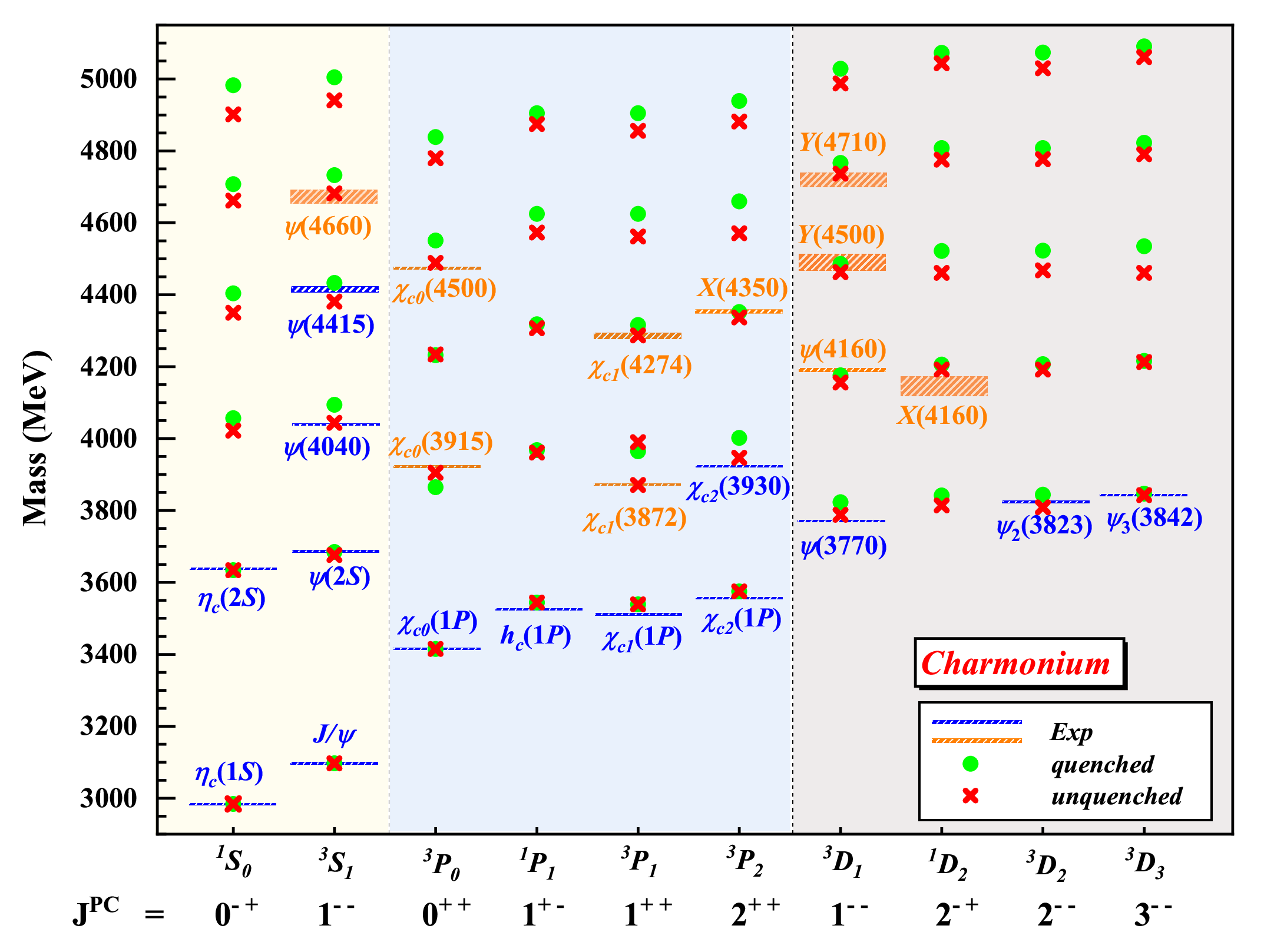}
\caption{
The charmonim spectrum calculated using the unquenched and quenched models, and the comparsion to the experimental measurements.
}
\label{fig:ccbar}
\end{center}
\end{figure*}

Ref.~\cite{Deng:2023mza} investigated the charmonium spectrum in the framework of an unquenched quark model incorporating coupled-channel effects.
In their numerical calculations, all open charmed-meson channels are taken into account using the once-subtracted approach, and a suppression factor is introduced to soften the hard vertices predicted by the $^3P_0$ model at high momentum transfer.
This framework provides a good simultaneous description of both the masses and decay widths of the well-established charmonium states.
In addition, Ref.~\cite{Deng:2023mza} presents predictions for higher-mass S-, P-, and D-wave charmonium states up to approximately 5.0 GeV.
Figure~\ref{fig:ccbar} shows the spectrum calculated within the unquenched model by Ref.~\cite{Deng:2023mza}, along with comparisons to the quenched model and experimental measurements.

\section{vector charmonium-like states at BESIII}

BESIII has measured a series of cross sections for $e^+e^-$ annihilation processes as functions of $\sqrt{s}$, such as $e^+e^-\to\pi\pi\psi$, $\pi\pi h_c$, $KK\psi$, $\eta\psi$, $\omega\chi_{c0,1,2}$, and $\phi\chi_{c0,1,2}$~\cite{fig3_1,fig3_2,fig3_3,fig3_4,fig3_5,fig3_6,fig3_7,fig3_8,fig3_9,fig3_10}. The line shapes of these cross sections frequently exhibit nontrivial structures, and it is a reasonable assumption that conventional vector charmonium states are present among them.
By analyzing measurements of each process, several structures have been reported.
There are four main structures above 4.4 GeV: $\psi(4415)$, $\psi(4500)$, $\psi(4660)$, and $\psi(4710)$.
The structure $\psi(4415)$ around 4.415 GeV appears in many processes, e.g., $D_s^* D_s^*$~\cite{BESIII:2023wsc}.
$\psi(4500)$ and $\psi(4710)$ are observed in the $K^+K^-J/\psi$~\cite{BESIII:2023wqy} and $K_S^0 K_S^0 J/\psi$~\cite{BESIII:2022joj} channels.
$\psi(4660)$ is also well established in the $\pi\pi\psi(2S)$~\cite{Belle:2014wyt,BESIII:2021njb} decay.
In addition, similar structures with masses around 4.6 to 4.75 GeV are present in the line shapes of
$e^+e^-\to D_s^{+}D_{s1}^{*-}(2536)$, $D_s^{+}D_{s2}^{*-}(2573)$~\cite{BESIII:2024qfi}, $\phi\chi_{c1,2}$~\cite{BESIII:2024qfi}, and $K^+K^-\psi(2S)$~\cite{BESIII:2024vjf}.

However, the tricky issue is that the structures observed in these processes are not fully consistent, and it is difficult to determine whether they correspond to the same state owing to the limited measurement precision. The vector conventional charmonium states with the quarntum number of $J^{PC}=1^{--}$ above 4.4 GeV include $\psi(4S)$, $\psi(5S)$, $\psi(3D)$, and $\psi(4D)$ as shown in Figure~\ref{fig:ccbar}, which could be the candidates of these structures observed in experiments.

As discussed in Ref.~\cite{Deng:2023mza}, the vector resonances $\psi(4415)$ and $\psi(4660)$ show a preference for the $\psi(4S)$ and $\psi(5S)$ assignments, respectively. Meanwhile, the newly discovered vector states $Y(4500)$ and $Y(4710)$ by the BESIII experiment are more consistent with the $\psi(3D)$ and $\psi(4D)$ interpretations, despite the fact that the $\psi(5S)$ assignment for $Y(4710)$ remains a viable option.
In addition, Ref.~\cite{Deng:2023mza} has calculated the partial decay widths of $\psi(5S)$, $\psi(3D)$, and $\psi(4D)$ into the open-charm channels $D^{*+}_s D^{*-}_s$, $D^{+}_s D^{*-}_{s1}(2536)$, and $D^{+}_s D^{*-}_{s1}(2573)$.

\section{Fit to $\sqrt{s}$-dependent cross sections}

We perform a simultaneous $\chi^2$-minimized fit to the line shapes of the processes $e^+e^-\to D_s^{+}D_{s1}^{*-}(2536)$, $D_s^{+}D_{s2}^{*-}(2573)$, $\phi\chi_{c1,2}$, $K^+K^-J/\psi$, $K_S^0 K_S^0 J/\psi$, and $K^+K^-\psi(2S)$.
Four resonant structures, $\psi(4230)$, $\psi(4500)$, $\psi(4660)$, and $\psi(4710)$, are employed to describe these datasets.

The $\sqrt{s}$-dependent cross sections are described with a coherent sum of the 
relativistic Breit-Wigner functions which is defined as

\[
BW(\sqrt{s}) = \frac{\sqrt{\Gamma^{\text{tot}}}}{s - M^2 + iM\Gamma^{\text{tot}}} \frac{\sqrt{\Phi(\sqrt{s})}}{\sqrt{\Phi(M)}},
\]
where
\[
\Phi(\sqrt{s}) = \frac{q^{2l+1}(\sqrt{s})}{s},
\]
$l$ denotes the orbital angular momentum, which takes the value zero for all processes except the $D^{+}_s D^{*-}_{s2}(2573)$ channel, where $l=2$ corresponding to a D-wave between $D^{+}_s$ and $D^{*-}_{s2}(2573)$.
$\Phi(\sqrt{s})$ is the phase space factor.

\begin{figure}[htbp]
\begin{center}
\includegraphics[scale=0.9]{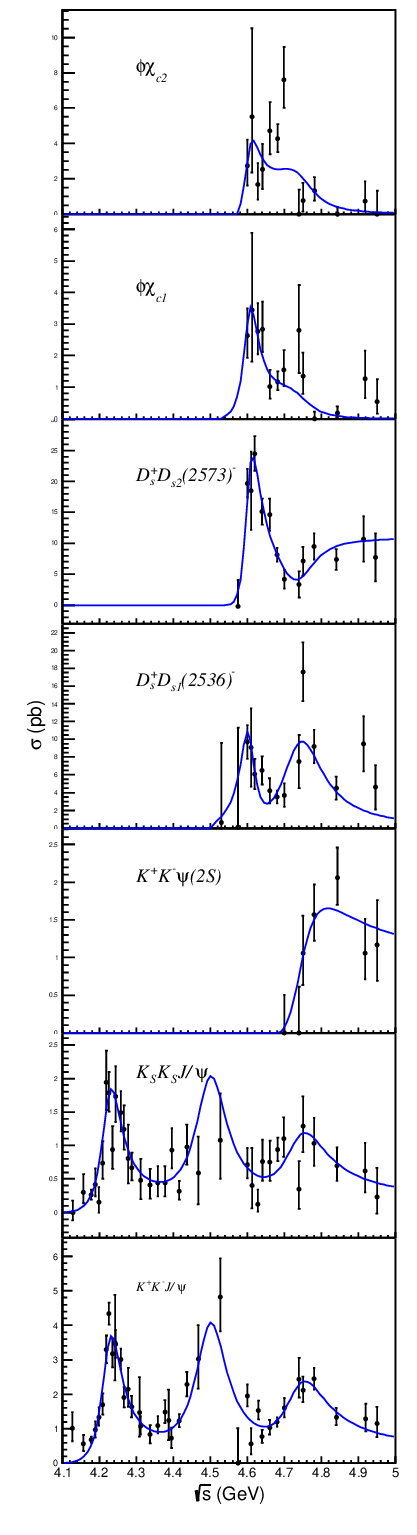}
\caption{
The $\sqrt{s}$-dependent cross sections and  fit results.
}
\label{fig:crs_fit}
\end{center}
\end{figure}

In the fit, the resonant components employed in each channel are listed in Table~\ref{res_comp}. Briefly, contributions from $\psi(4230)$, $\psi(4500)$ and $\psi(4710)$ are considered in the $K^+K^-J/\psi$ and $K_S^0 K_S^0 J/\psi$ channels, and the phase configurations for these two channels are identical. The amplitude ratio of each resonance in $K_S^0 K_S^0 J/\psi$ relative to $K^+K^-J/\psi$ is constrained to 0.5 under the assumption of isospin symmetry. For the remaining channels, $\psi(4660)$ and $\psi(4710)$ are included with independent phase parameters.
The $\chi^2$ is defined as 
\begin{equation}
\chi^2 = \sum_{i=1}^{7}\sum_{j=1}^{N_i} (\frac{x_{ij}-\mu_{ij}}{\sigma_{ij}})^2
\end{equation}
where $i$ takes values from 1 to 7, corresponding 
to the measurements of the seven channels. 
$x_{ij}$ and $\sigma_{ij}$ are the measured cross sections and 
corresponding statistical errors in each energy points as shown in Figure~\ref{fig:crs_fit}, 
$\mu_{ij}$ is the expected value of 
the cross section at each energy
point calculated with the fitting shape mentioned above, and
$N_i$ is the number of energy points in each measurement.
The systematic uncertainty of each measurement is not considered in the fit, since the values at different energy points for each channel are highly correlated with each other and do not significantly affect the line shape, but only the overall amplitude.
The fit results are shown in Figure~\ref{fig:crs_fit}, and the yield parameters are listed in Table~\ref{res_comp}. The fit quality with $\chi^2/\mathrm{ndf}=155.4/118$ indicates a good description of the data.
When fitting with a coherent sum of several Breit-Wigner functions, multiple solution conditions arise due to interference between these functions. These solutions share the common masses and widths of the resonances but differ in amplitude magnitudes. Here, we focus solely on the extraction of resonant parameters and neglect the amplitude part; thus, the multiple solutions do not affect our conclusions, and only one solution is presented in this article.

According to the fit results, based on the current measurement precision, three resonances are sufficient to describe the line shapes of the processes $e^+e^- \to K^+K^-J/\psi$ and $e^+e^- \to K_S^0 K_S^0 J/\psi$. Although both $\psi(4660)$ and $\psi(4710)$ are included in the $K^+K^-\psi(2S)$ channel, the fit results indicate that the dominant contribution arises from $\psi(4710)$, while the contribution from $\psi(4660)$ is negligible.

It is worth noting that we did not include the $e^+e^-\to D_s^{*+}D_s^{*-}$ channel in the fit, although its measurement has high precision. We attempted to incorporate this channel into the fit but were unsuccessful. None of the resonances employed in the fit, including the $\psi(4230)$, can describe the potential resonant contributions to the $D_s^{*+}D_s^{*-}$ process. We are still investigating the $D_s^{*+}D_s^{*-}$ data and hope to find a resolution in the future by introducing additional resonance contributions.

\begin{table}
\caption{
Coherent sum of the relativistic Breit-Wigner function for each channel.
$BW_{1,2,3,4}$ are corresponding the 
$\psi(4230)$, $\psi(4500)$,  $\psi(4660)$, and $\psi(4710)$,
$a_{ij}$ and $\phi_{ij}$ are the coefficients and the relative phase of the resonances.
}
\begin{tabular}{p{2cm}p{4cm}p{8cm}}
\hline\hline
 Index&Channels &  Shape\\
\hline
1&$K^+K^-J/\psi$&$a_{11}\cdot BW_1\cdot e^{i\phi_{11}}+a_{12}\cdot BW_2\cdot e^{i\phi_{12}}+a_{13}\cdot BW_3$\\
2&$K_sK_sJ/\psi$&$0.5\cdot a_{11}BW_1\cdot e^{i\phi_{11}}+0.5\cdot a_{12}BW_2\cdot e^{i\phi_{12}}+0.5\cdot a_{13}BW_3$\\
3&$K^+K^-\psi(2S)$&$a_{33}\cdot BW_3\cdot e^{i\phi_{33}}+a_{34}\cdot BW_4$\\
4&$D_s^{*+}D_{s1}(2536)^-$&$a_{43}\cdot BW_3\cdot e^{i\phi_{43}}+a_{44}\cdot BW_4$\\
5&$D_s^{*+}D_{s2}(2573)^-$&$a_{53}\cdot BW_3\cdot e^{i\phi_{53}}+a_{54}\cdot BW_4$\\
6&$\phi\chi_{c1}$&$a_{63}\cdot BW_3\cdot e^{i\phi_{63}}+a_{64}\cdot BW_4$\\
7&$\phi\chi_{c2}$&$a_{73}\cdot BW_3\cdot e^{i\phi_{73}}+a_{74}\cdot BW_4$\\
\hline\hline
\label{res_comp}
\end{tabular}
\end{table}

\begin{table}
\caption{
The fit results include the masses and widths of the resonances.
}
\begin{tabular}{p{3cm}p{5cm}}
\hline\hline
 Structures &  Parameters (MeV)\\
\hline
$\psi(4230)$&$M_1=4225.39\pm1.95$\\
&$\Gamma_1=66.31\pm5.04$\\
\hline
$\psi(4500)$&$M_2=4496.77\pm12.19$\\
&$\Gamma_2=104.28\pm23.80$\\
\hline
$\psi(4660)$&$M_3=4604.03\pm2.92$\\
&$\Gamma_3=48.48\pm5.02$\\
\hline
$\psi(4710)$&$M_4=4736.38\pm7.03$\\
&$\Gamma_4=137.25\pm20.11$\\
\hline\hline
\label{sum_sys}
\end{tabular}
\end{table}

\section{Discussion}
We describe the $\sqrt{s}$-dependent line shapes of the $e^+e^-\to D_s^{+}D_{s1}^{*-}(2536)$, $D_s^{+}D_{s2}^{*-}(2573)$, $\phi\chi_{c1,2}$, $K^+K^-J/\psi$, $K_S^0 K_S^0 J/\psi$, and $K^+K^-\psi(2S)$ cross sections measured by the BESIII experiment using the four resonant structures $\psi(4230)$, $\psi(4500)$, $\psi(4660)$, and $\psi(4710)$, by performing a simultaneous $\chi^2$-minimized fit.
We find that the resonant structures are sufficient, and the contributions from the continuum process are insignificant.
The processes $e^+e^-\to D_s^{+}D_{s1}^{*-}(2536)$, $e^+e^-\to D_s^{+}D_{s2}^{*-}(2573)$, and $e^+e^-\to \phi\chi_{c1,2}$ are all dominantly produced via the $\psi(4660)$ and $\psi(4710)$ decays.

From the fit, we extract the masses and widths of these resonances. There are four vector charmonium states with the quantum number $1^{--}$ above 4.4 GeV, namely $\psi(4S)$, $\psi(3D)$, $\psi(5S)$, and $\psi(4D)$, which could be candidates for the observed structures in the data. However, there is still a discrepancy between their measured masses and the theoretical predictions, which makes it difficult to assign these conventional charmonium states to the observed structures with a high confidence level.

More high-precision measurements, especially for open-charm processes which often have corresponding theoretical predictions, are essential to clarify these ambiguities.
Several measurements for two- and three-body final-state channels have already been reported by BESIII, and these processes will be incorporated into our analysis framework.

\bibliographystyle{unsrt}

\end{document}